\def\3he{{$^3${\rm He}}}

\def\eg{{\it e.g.,\ }}
\def\ie{{\it i.e.,\ }}
\def\etal{{\it et al.}}

\def\slD{\raise.15ex\hbox{$/$}\kern-.53em\hbox{$D$}}
\def\slA{\raise.15ex\hbox{$/$}\kern-.53em\hbox{$A$}}
\def\dsl{\raise.15ex\hbox{$/$}\kern-.57em\hbox{$\Delta$}}
\def\slp{{\raise.15ex\hbox{$/$}\kern-.57em\hbox{$\partial$}}}
\def\nsl{\raise.15ex\hbox{$/$}\kern-.57em\hbox{$\nabla$}}
\def\sla{\raise.15ex\hbox{$/$}\kern-.57em\hbox{$\rightarrow$}}
\def\slla{\raise.15ex\hbox{$/$}\kern-.57em\hbox{$\lambda$}}
\def\slb{\raise.15ex\hbox{$/$}\kern-.57em\hbox{$b$}}
\def\slr{\raise.15ex\hbox{$/$}\kern-.57em\hbox{$r$}}
\def\lnp{\raise.15ex\hbox{$/$}\kern-.57em\hbox{$p$}}
\def\lnk{\raise.15ex\hbox{$/$}\kern-.57em\hbox{$k$}}
\def\lnK{\raise.15ex\hbox{$/$}\kern-.57em\hbox{$K$}}
\def\lnq{\raise.15ex\hbox{$/$}\kern-.57em\hbox{$q$}}
\def\nna{\raise.15ex\hbox{$/$}\kern-.57em\hbox{$a$}}

\def\a{\alpha}
\def\be{{\beta}}

\def\munu{{\mu\nu}}

\def\cD{{\cal D}}

\def\cH{{\cal H}}

\def\cN{{\cal N}}

\def\bbR{{I\kern-0.3em R}}

\def\fulltriangle{{{{{{{{{\pmb{\triangle}\kern-.65em\bullet}\kern-.4em
{\raise1.2ex\hbox{.}}}
\kern-.4em{\raise1.0ex\hbox{.}}}\kern-.2em{\raise1.0ex\hbox{.}}}
\kern-.4em{\raise.1ex\hbox{.}}}\kern-.4em{\raise.2ex\hbox{.}}}
\kern-.2em{\raise.35ex\hbox{.}}}\kern.1em{\raise.2ex\hbox{.}} }}

\def\hexagon{{\tenpoint
\langle\kern-.1em{\raise.2cm\hbox{$\overline{\hskip.7em\relax}$}}
\kern-.7em{\lower.3ex\hbox{$\underline{\hskip.7em\relax}$}}\kern-.075em
 \rangle}}

\def\pentagon{{\tenpoint
\raise.5ex\hbox{$\widehat{\qquad}$}\kern-1.8em{\backslash
\kern-.1em{\lower.3ex\hbox{$\underline{\kern.75em}$}}\kern-.05em/} }}

\def\pmb#1{\setbox0=\hbox{$#1$}%
\kern-.025em\copy0\kern-\wd0
\kern.05em\copy0\kern-\wd0
\kern-.025em\raise.0433em\box0 }

\def\q2{{Q^2}}
\def\gtwid{\raise.3ex\hbox{$>$\kern-.75em\lower1ex\hbox{$\sim$}}}
\def\ltwid{\raise.3ex\hbox{$<$\kern-.75em\lower1ex\hbox{$\sim$}}}
\def\12{{1\over2}}

\def\part{\partial}

\def\low#1{\lower.5ex\hbox{${}_#1$}}

\def\partt{\raise.15ex\hbox{$\widetilde$}{\kern-.37em\hbox{$\partial$}}}

\def\vev#1{{\left\langle{#1}\right\rangle}}

\def\topppageno1{\global\footline={\hfil}\global\headline
={\ifnum\pageno<\firstpageno{\hfil}\else{\hss\twelverm --\ \folio
\ --\hss}\fi}}

\def\toppageno2{\global\footline={\hfil}\global\headline
={\ifnum\pageno<\firstpageno{\hfil}\else{\rightline{\hfill\hfill
\twelverm \ \folio
\ \hss}}\fi}}

\def\boxit#1{\vbox{\hrule\hbox{\vrule\kern3pt
  \vbox{\kern3pt#1\kern3pt}\kern3pt\vrule}\hrule}}

\documentstyle[preprint,revtex]{aps}
\begin{document}
\def\tilda{{\tilde\alpha}}
\def\nbar{{\bar n}}
\def\underk{{\underline{k}}}
\def\underr{{\underline{r}}}
\def\underx{{\underline{x}}}
\def\undery{{\underline{y}}}
\def\underz{{\underline{z}}}
\preprint{BROWN-HET-849}
\preprint{August 1992}
\medskip
\draft
\begin{title}
The Entropy of the Gravitational Field
\end{title}
\vskip .2in
\author{R. Brandenberger, T. Prokopec}
\begin{instit}
Department of Physics, Brown University, Providence, Rhode Island
02912, USA
\end{instit}
\smallskip
\centerline{and}
\smallskip
\author{V. Mukhanov\cite{slava}}
\begin{instit}
Institute of Theoretical Physics, ETH H\"onggerberg, CH-8093 Z\"urich,
Switzerland
\end{instit}
\vskip 0.5in
\centerline{ABSTRACT}
\begin{abstract}
We derive a formula for the nonequilibrium entropy of a classical
stochastic field in terms of correlation functions of this field.  The
formalism is then applied to define the entropy of gravitational
perturbations (both gravitational waves and density fluctuations).  We
calculate this entropy in a specific cosmological model (the
inflationary Universe) and find that on scales of interest in
cosmology the entropy in both density perturbations and gravitational
waves exceeds the entropy of statistical fluctuations of the microwave
background.

The nonequilibrium entropy discussed here is a measure of loss
of information about the system.  We discuss the origin of the entropy in
our cosmological models and compare the definition of entropy in terms
of correlation functions with the microcanonical definition in quantum
statistical mechanics.
\end{abstract}
\newpage
\section{INTRODUCTION}

The concept of entropy plays an important role in all branches of physics.
Hence, it is no surprise that entropy can also be defined in general relativity
and cosmology. The most famous application of entropy in general relativity is
to black holes.
Bekenstein's [1] realization that the area of a black hole behaves like
an entropy led to the discovery of black hole radiation [2].
Other important applications of entropy in cosmology relate to the
characterization of the initial state of the universe [3] and to the
study of the post-bounce state in bouncing cosmologies.

Entropy expresses the loss of information about the system under
consideration [4]. Quite a long time ago, Jaynes [4] argued that
entropy expresses the extent of human ignorance about a system and is
therefore an antropomorphic concept. One can uniquely define entropy
only after having specified the position of the observer with respect
to the system.

Once we accept that entropy measures the loss of information about a
system, it becomes important to ask whether there is a natural
``coarse graining", i.e. can it be uniquely specified which part of
the information about a system is lost? The problem of defining the
entropy becomes the task of determining the correct coarse graining.

We wish to address this question in order to define the entropy of the
gravitational field. From a naive point of view, gravitational
instability, which
is responsible for the formation of structure in the
Universe, should lead to an increasing entropy - in agreement with
the second law of thermodynamics.
In the initial
state in which the gravitational field is (almost) uniform, the
gravitational entropy (almost) vanishes.
In contrast, the later state of the
gravitational field which results from gravitational instability can
be viewed as
a particular realization of some stochastic process producing density
perturbations and gravitational waves. Hence, there should be an associated
entropy which characterizes the naturalness of the occurrence of the given
distribution or, more quantitatively, measures the probability of the
distribution.

To characterize the measure of a state of gravitational radiation or density
perturbations in a quantitative manner, we need a well defined and well
justified notion of entropy of gravitational perturbations. Since the
states we
are interested in are far from thermal equilibrium, our task will be to define
an entropy for nonequilibrium systems in cosmology.

There exists already a considerable body of work on the definition
of nonequilibrium entropy, in particular in the context of cosmology. For
example, Smolin [5] derived a formula for the entropy of a ``quantum
field" state of gravitational radiation and applied it to estimate the entropy
of some astrophysical sources of gravitational radiation.

In a series of paper, Hu and Paron [6], Kandrup [7,8] and Kandrup
and Hu [9] have discussed the entropy of particles produced in an
expanding universe. They gave a definition of entropy based on the single
particle distribution function (density matrix) and showed that this entropy
increases in time (if the initial state is free of correlations) as a result of
particle production. This definition of entropy (and in particular the role of
coarse graining and loss of correlations) was further discussed by Habib and
Kandrup [10].

Some interesting speculations about the entropy $S_g$ of the gravitational
field have been made by Penrose [3] and Hu [11],  who propose that
$S_g$ is proportional to the integral of the Weyl tensor squared $C^2$ over
space. This definition expresses the expectation that metric fluctuations
should give rise to entropy. With this formulation, Penrose's initial condition
criterion [3] $C=0$ for the universe is equivalent to the assumption that
the universe starts in a state of vanishing gravitational entropy.

In this paper we use two quite different approaches to define the
nonequilibrium entropy of cosmological perturbations. One of them is based on
the microcanonical ensemble [12] of quantized gravitational
perturbations while the other is a formula for entropy which
can be associated with the stochastic distribution which describes the
state of the classical gravitational field.  We show that
these two definitions are in agreement and discuss the physical meaning of the
entropy of gravitational fluctuations. Then, we apply our definitions to
estimate the entropy of gravitational waves and linear density inhomogeneities
produced by inflation [13].  We also indicate how to apply our methods to
other cosmological models, \eg those based on phase transitions [14].

Our analysis is based on the fact that the theory of gravitational waves and
of linearized density perturbations in an expanding universe can be
reduced [15] to the study of a real scalar field in an external classical
background. Hence, we will investigate the more general question of how to
define the entropy of a scalar field in a nonequilibrium state. We can either
study the quantum theory of this field and use the microcanonical ensemble, or
we can view the classical field as a stochastic process and determine its
entropy.

The paper is organized in the following manner: after some general comments
about nonequilibrium entropy in section 2, we give in sections 3 and 4 the
quantum and classical definitions of entropy. In section 5 we show the
equivalence of the two definitions when applied to situations when both are
applicable. We also demonstrate that the entropy is a result of coarse
graining.  In section 6 we briefly review the
gauge invariant theory of cosmological perturbations upon which our definition
of the entropy of the gravitational field is based. For a pedagogical
introduction, the reader is referred to Ref.~16, for an extensive review to
Ref.~15. Section 7 contains the main applications of our work: the
evaluation
of the entropy of gravitational waves and density perturbations in inflationary
universe models.

We use units in which $\hbar=k_B=c=1$. Greek indices run over space-time
variables, Latin indices only over spatial variables.

\section{NONEQUILIBRIUM ENTROPY}

We will first develop a general definition of entropy for a system far from
thermal equilibrium, based on the microcanonical ensemble [12]. Let us
assume that the state of some physical system can be completely described by a
set of discrete variable $J=\left\{I,i,j,\dots\right\}$. If we know that the
system is in a certain state $J$, then the information about the state of the
system is complete and hence the entropy  should be zero, as follows from the
general definition of entropy in information theory according to which entropy
means the loss of information. If, on the other hand, we only know the
probability distribution $P_J$ for the system, $P_J$ being the probability to
find the system in state $J$, then the associated entropy is [4,12]
\begin{equation}
S=-\sum_J P_J\ln P_J
\end{equation}

Now let us assume that we are not interested in or cannot measure the complete
(fine grained) state of the system, but only some coarse grained
characteristics, \eg the value of the variable $I$. The coarse grained state
can be described by a distribution function $P_I$, and the associated entropy
is given by the analog of (1) where we sum only over the index $I$. This
entropy characterizes the measure of solutions of the dynamical system which
leads to the particular coarse grained state.

If the variable  $J$ in (1) is continuous, some complications arise.
If $\cD J$ is the measure on the space of state, then a probability density
$p(J)$ can be defined by
\begin{equation}
\cD P(J)=p(J)\cD J,
\end{equation}
where $\cD P(J)$ is the probability to find the system in the volume $\cD J$
around the state $J$. For example, in a system of $n$ particles with
Cartesian coordinates $x_1,\dots x_n$ and momenta $p_1,\dots p_n$, we would
have $J\equiv(x_1 \dots , x_n,p_1,\dots,p_n)$ and $\cD
J=\prod^n_{i=1}(dx_i dp^i)$. In a
system with an infinite number of particles, $\cD J$ becomes a measure in the
space of functionals.

To derive the formula for the entropy of a system with continuous $J$ starting
from Eq.~(1), we first divide phase space into sufficiently small cells
$J_1,J_2\dots$
with volume elements $\Delta J_1,\Delta J_2,\dots$. The probability to find the
system in cell $n$ is
\begin{equation}
P_{J_n} \simeq p(J_n)\Delta J_n,
\end{equation}
and hence from (1) the entropy is
\begin{equation}
S=-\sum_n P_{J_n}\ln P_{J_n} \simeq - \sum_n p(J_n) (\ln p(J_n)) \, \Delta J_n
-
   \sum_n p(J_n) \ln(\Delta J_n) \Delta J_n.
\end{equation}
This expression has no limit for $\Delta J_n\to0$ because of the diverging
second term which in general depends on $\{ p (J_n) \}$.  This terms
represents the information about the process of coarse graining.

However, for a simple coarse graining with $\Delta J_1=\Delta J_2=
\dots$, the second term in (4) does not depend on the probability distribution
$P(J)$ and can hence be neglected as some irrelevant additive constant
$-\ln\Delta J$ to the entropy. Note that in a quantum dynamical system, there
is a natural choice $\Delta J= (2\pi\hbar)^n$ due to the uncertainty
principle. We
conclude that in the case of a continuous probability distribution, the entropy
is defined by coarse graining and depends on the measure in the phase space of
the system. Dropping the second term in (4) and taking the continuum limit
$\Delta J\to0$ gives
\begin{equation}
S=-\int p(J) \ln p(J) \cD J,
\end{equation}
where $\cD J$ is the functional measure for the variable $J$.

\section{MICROCANONICAL DEFINITION OF ENTROPY \\
 FOR A QUANTIZED FIELD}

Let us return to a system whose phase space is described by a set of
discrete variables. Furthermore,
we consider the case when the entire system consists of
$\cN$ identical subsystems (\eg $\cN$ photons), each
characterized by a discrete set of variable $\left\{I,i,j,\dots\right\}$. We
assume that there is some principal quantum number $I$ which is completely
distinguishable, \ie any two states with different $I$ can be
experimentally
distinguished, and that the other numbers $i,j,\dots$ correspond to different
but experimentally indistinguishable   states with the same value of $I$. As an
example, for a gas of photons in a box we can take $I$ to be the energy of a
photon, and $i,j,\dots$ to correspond to different directions of motion.

The source of entropy in the above setup is the loss of information coming from
the indistinguishability of states with identical $I$ but different
$i,j, \dots$  States labelled by $I$ can be assigned a degeneracy
$g_I$ which equals the number of microphysical states with identical quantum
number $I$.

Let us now assume that $\nbar_I$ subsystems have the same principal quantum
number $I$. For the moment we assume that the spectrum of the system
$\left\{\nbar_I\right\}$, \ie the number $\nbar_I$ of subsystems  with
principal
quantum number $I$ (for all $I$), is fixed. Our goal is to calculate the number
of possible microphysical states with a given spectrum which are in principle
distinguishable. The calculation is done for systems with Bose statistics, \eg
photons, gravitons or scalar type cosmological perturbations.

The problem reduces to calculating the number of possible and distinguishable
ways in which $\nbar_I$ subsystems can be distributed among $g_I$ cells. This
number is
\begin{equation}
W_{\nbar_I} = {(g_I -1+\nbar_I)!\over \nbar_I!(g_I-1)!} .
\end{equation}
To obtain this expression, note that there are $(\nbar_I+g_I-1)!$ ways of
dividing $\nbar_I$ objects by $g_I-1$ cell divisions. However, both the
particles and the cell divisions are  indistinguishable and hence we must
divide by $\nbar_I!(g_I-1)!$.

For a system of $\cN$ subsystems with spectrum $\left\{\nbar_I\right\}$
(obeying $\sum_I\nbar_I=\cN$), the phase volume (number of possible states)
will be
\begin{equation}
\Gamma_{\{\nbar_I\}}=\prod_I W_{\nbar_I} .
\end{equation}

The next step is to assume that all possible states are equally probable. In
this case, the probability for any state $\a$ with the
given spectrum $\{\nbar_I\}$ is
\begin{equation}
P_{\{\nbar_I\}} (\a)={1\over\Gamma_{\{\nbar_I\}}} .
\end{equation}
{}From (1) it follows that the corresponding entropy of the system with
definite spectrum $\{\nbar_I\}$ is
\begin{equation}
S=-\sum_\a P_{\{\nbar_I\}}(\a)\ln P_{\{\nbar_I\}}(\a)=\ln
        \Gamma_{\{\nbar_I\}}=\sum_I\ln W_{{\nbar_I}},
\end{equation}
taking into account the normalization condition
\begin{equation}
\sum_\a P_{\{\nbar_I\}}(\a)=1.
\end{equation}

If $\nbar_I\gg1$, then Stirling's formula can be applied to approximate
$W_{\nbar_I}$ in (9). In this case, the entropy becomes
\begin{equation}
S=\sum_Ig_I   \left[(n_I+1) \ln (n_I+1)-n_I\ln n_I\right]
\end{equation}
where $n_I=\nbar_I/g_I$ are the occupation numbers.

All that was assumed in the above considerations is that the spectrum
$\{\nbar_I\}$ is well defined. At no point was thermodynamic equilibrium
invoked. Hence, (11) gives a formula for the entropy of a statistical system
with definite spectrum which is valid both in and far out of thermodynamical
equilibrium. It is based on the microcanonical ensemble.

The simplest application of this formula for the entropy is to a black body
spectrum of photons with
\begin{equation}
n_k={2\over e^{\be k}-1}
\end{equation}
with $\be=1/T$ and $k=|k|$. In this case, each photon is a subsystem. The
principal quantum number $I$ is the energy $k$, and the other quantum numbers
$i,j,\dots$ correspond to the directions of photon propagation. The degeneracy
$g_k$ of level $I=k$ is
\begin{equation}
g_k={4\pi\over3}Vk^2dk
\end{equation}
where $V$ is the volume of space. Substituting (12) and (13) into (11) we
obtain the entropy density of the black body background
\begin{equation}
s={S\over V}\sim {4\pi\over3}T^3 \, .
\end{equation}

In the above example, the origin of the entropy is the absence of information
about the direction of propagation of the photons.

If the spectrum $\{\nbar_I\}$ is not well known, there is an additional source
of entropy.  Let us assume a probability distribution
$P(\{\nbar_I\})$ for different spectra. Note that even the total number $\cN$
of subsystems need not be fixed. In this case, the space of all possible states
is the direct sum of states for the different spectra $\{\nbar_I\}$. The
probability to find the system in a state $\a_{\{\nbar_I\}}$ with spectrum
$\{\nbar_I\}$ is
\begin{equation}
P(\a_{\{\nbar_I\}})=P(\{\nbar_I\}){1\over\Gamma(\{\nbar_I\})} \, ,
\end{equation}
and (from (1)) the entropy will be
\begin{eqnarray}
S &=& -\sum_{\{\nbar_I\}} \sum_{\a{\{\nbar_I\}}} P(\a_{\{\nbar_I\}}) \ln
    P(\a_{\{\nbar_I\}}) \nonumber \\
  &=& \sum_{\{\nbar_I\}} P(\{\nbar_I\}) \ln \Gamma (\{\nbar_i\})
-\sum_{\{\nbar_I\}} P(\{\nbar_I\})\ln P(\{\nbar_I\})
\end{eqnarray}
where $\sum_{\{\nbar_i\}}$ stand for summation over all possible spectra
$\{\nbar_I\}$. If the spectrum is completely specified, then
$P(\{\nbar_I\})=1(0)$ for $\{\nbar_I\}=\{\nbar_I^0\}(\{\nbar_I\}\neq
\{\nbar_I^0\})$ and (16) reduces to (9). In the general case,
there are two contributions to the entropy. The first term in (16) is due to
the absence of information about the non-principal quantum numbers of the
system for a fixed spectrum, the second comes from our ignorance of the precise
spectrum.

In the case of cosmological perturbations, the distribution function
$P(\{\nbar_I\})$ for the spectrum is well localized at a particular spectrum
$\{\nbar_I^0\}$ and hence
\begin{equation}
S\simeq\ln \Gamma (\{\nbar_I^0\}) \, .
\end{equation}

However, there are examples where the second term in (16) gives the main
contribution to the entropy. For example, in the case of a black hole it is
impossible to have information about the spectrum of the configuration making
up the hole, since this information is hidden behind the horizon. If $W$ is the
number of possible different spectra for a black hole of fixed mass, and if we
assume that all of these spectra have equal probability, then, neglecting the
first term in (16), we get
\begin{equation}
S\simeq\ln W \, .
\end{equation}
If the black hole is quantized, then $W$ is finite. In fact, counting the
number of spectra of a black hole with fixed mass gives a formula for the
entropy in agreement with the classical result [17].

Let us return to the discussion of the formula (11) for the entropy of a
quantum system with fixed spectrum. In the classical limit $n_I\gg1$, the
equation simplifies to
\begin{equation}
S\simeq\sum_I g_I\ln n_I \, .
\end{equation}
In order to apply this formula, the notion of particles (which is required to
be able to determine $n_I$) must be well defined. However, when considering
quantum fields in some external field (\eg cosmological perturbations in an
expanding background space-time), the notion of particles is not always well
defined. It is hence desirable to have a formalism which generalizes the
definition of entropy to situations where the number representation is not well
defined.

For large occupation numbers $n_I$, the classical limit should give a good
description of the dynamics of the system. It is therefore convenient to derive
a formula for the entropy directly in terms of the classical field. In the
cases when occupation numbers can be defined for this field, the new definition
should reduce to (19).

In the next section, we will give a definition of entropy of a classical field
based on the theory of stochastic processes and show that in the region where
both definitions of entropy are applicable they agree.

\section{ENTROPY OF A CLASSICAL FIELD}

Formulas (11) and (19) for the entropy of a field are only applicable if the
spectrum of occupation numbers is well defined. In order for this to be the
case, there must be a well defined notion of particles for the field under
consideration.

As will be shown in Section 6, the description of cosmological perturbations
and gravitational waves in the Universe can be reduced to the study of a scalar
field $\varphi (\underline x , t)$ with a time dependent effective mass:
[15,16]
\begin{equation}
\varphi^{\prime\prime}-c_s^2\Delta \varphi -{z^{\prime\prime}\over z}
\varphi =0 \, .
\end{equation}
Here, $c_s$ is the speed of propagation of perturbations. For
gravitational waves $c_s=1$, whereas for cosmological perturbations in a
Universe with hydrodynamical matter $c_s$ is the speed of sound. The time
dependent function $z$ depends on the system and on the background. For
gravitational waves $z=a$, whereas $z$ is a complicated function of the
background parameters in the case of cosmological perturbations (for details
see Ref.~15 and Section 6). In the above a prime denotes
differentiation time $\eta$.

The quantization of a scalar field obeying (20) is equivalent to the
quantization of a scalar field in some external classical field. If
$z^{\prime\prime}/z=0$ and $c_s^2={\rm const}\neq0$, then there is no coupling
and the quantum theory for this field (in particular the notion of particle) is
well defined. As will be shown in Section 6, this applies for both
cosmological perturbations and gravitational waves in a radiation dominated
Universe. However, this is not the general case. The ratio $z^{\prime\prime}/z$
can be nonvanishing and $c_s^2$ may be zero. For example, for cosmological
perturbations in a matter dominated Universe $c_s=0$. In this case the
solutions of (20) do not have oscillatory character and it is not possible to
define the notion of particles. Hence it is not possible to define occupation
numbers and Equations (11) and (19) are inapplicable.

However, if the perturbations are sufficiently large, the scalar field
$\varphi$
can be treated classically (in the case when occupation numbers {\it are}
defined, the condition for classicality is $n_I\gg1$). In order to
define a notion of entropy valid in this case, we address the more general
question of defining the entropy of a classical scalar field with an action
which is quadratic in field variable and canonical momentum (see also
Ref.~18).

The source of the entropy is in this case the ignorance about the exact
field configuration. A state of the system at some
time $t$ is specified by the values of the field $\varphi (\underline
x , t)$
and its canonical momentum $\pi(\underline x , t)$ at all points $\underline
x$ in space. We assume that all we know is the probability distribution
$P(\varphi (\underline x),\pi(\underline x))$ of the field and its canonical
momentum,
\ie we view $\varphi (\underline x)$ as a stochastic classical field.

The above situation is realized in many situations of interest in cosmology.
For example, in inflationary Universe models, the amplification of scalar field
fluctuations during the period of exponential expansion of the Universe and the
nontrivial transition of quantum fluctuations to classical ones leads to a
squeezed state [19,20] for the scalar field starting from the
vacuum state at the beginning of inflation. The Gaussian random state is
characterized by definite correlation functions $\vev{\varphi
(\underline x,t)\varphi (\underline y, t)}$,
$\vev{\pi(\underline x,t)\pi(\underline y, t)}$ and
$\vev{\varphi(\underline x,t)\pi(\underline y,t)}$, where $\vev{q}$ stands for
the ensemble average of the quantity $q$, (which coincides with the space
average of $q$ for a spatially homogeneous stochastic process).

If the initial state of the system is Gaussian, and if the Hamiltonian is
quadratic, then time evolution will preserve the Gaussian character of the
state. We will assume, as is the case for linear cosmological perturbations and
gravitational waves, that the state is Gaussian at all times. Therefore, the
probability distribution $P(\varphi (\underline x),\pi(\underline x))$ can be
expressed in terms of the above two-point correlation functions. Hence, also
the entropy of the system must be expressible in terms of two point correlation
functions.

In the following, we will derive a general expression for the entropy of a
stochastic Gaussian field in terms of its two-point correlation functions.

Starting point of the analysis is formula (5) for the
entropy in terms of the probability distribution. In our case, the continuous
variable $J$ stands for a point in phase space. To justify the
choice of the measure in (5), we divide phase space $(\varphi,\pi)$
at every point $\underline{x}$ in space
into units of volume $2 \pi \hbar$
(the smallest volume the fields can be localized
in by the uncertainty principle) and calculate the probability $\Delta P$ that
the fields lie in the bin $\Delta_J$:
\begin{equation}
\Delta P_J=\int_{\Delta_J} P(\varphi (\underline x),\pi(\underline x))\cD
\varphi (\underline x) \cD \pi(\underline x) \, .
\end{equation}
Here, the integration ranges over fields $\varphi (\underline x)$ and
$\pi(\underline x)$ which lie in bin $\Delta_J$ and $\cD \varphi
(\underline x)$
$\cD\pi(\underline x)$ denotes the functional integral measure for a
scalar field.

Thus, from the analysis of Section 2 we conclude that (apart from an irrelevant
constant), the entropy of the stochastic classical field is given by
\begin{equation}
S=-\int P(\varphi(\underline x),\pi(\underline x))\ln P(\varphi(\underline x),
\pi(\underline x)) \cD\varphi (\underline x) \cD \pi(\underline x) \, .
\end{equation}

For a Gaussian state, the probability distribution is
\begin{equation}
P(\varphi,\pi)={1\over W}\exp\left\{-{1\over2}\left(\varphi_xA^{xy}\varphi_y +
\pi_xB^{xy} \pi_y+2\varphi_x C^{xy} \pi_y\right)\right\}
\end{equation}
where the normalization factor $W$, determined by
\begin{equation}
\int P(\varphi,\pi)\cD \varphi\cD\pi=1 \, ,
\end{equation}
is given by
\begin{equation}
W=\int\exp\left\{-{1\over2}\left(\varphi_xA^{xy}\varphi_y+\pi_xB^{xy}
\pi_y+2\varphi_x
C^{xy}\pi_y\right)\right\}\cD\varphi\cD\pi \, .
\end{equation}
Here, we have used the short hand notation $\varphi(\underline x)=\varphi_x$
and
$A^{xy}=A(\underline x, \underline y)$, and the Einstein ``summation"
convention for repeated indices
\begin{equation}
\rho_xA^{xy}=\int d\underline x \rho(\underline x)A(\underline x, \underline y)
\end{equation}
is implied. For a homogeneous Gaussian state, the kernels $A^{xy}$, $B^{xy}$
and $C^{xy}$ depend only on the difference of the arguments
\begin{equation}
A^{xy}=A(\underline x,\underline y)=A(\underline x - \underline y)=
A(\underline y - \underline x)\, .
\end{equation}
The kernel of the operator $A^{-1}$ inverse to $A$ will be denoted by
$A^{-1}_{xy}$ (defined as usual by $A^{-1}_{xz} A^{zy} = \int d^3 z
A^{-1} (\underline{x} , \underline{z}) A (\underline{z} , \underline{y}) =
\delta^3 (\underline x - \underline y)$).
Note that the kernel $C$ does not in general vanish since in general
$\varphi$ and $\pi$ are not statistically independent.  A
nonvanishing $C$ reflects a correlation between $\varphi$ and $\pi$.

Our goal now is to express the kernels $A^{xy}$, $B^{xy}$ and $C^{xy}$ in terms
of the two point
correlation functions $\vev{\varphi(\underline x)\varphi(\underline
y)}$, $\vev{\pi(\underline x)\pi(\underline y)}$ and $\vev{\varphi(\underline
x)
\pi(\underline y)}$. As a first step, it is convenient to rewrite the
distribution function $P(\varphi, \pi)$ in terms of new variables which are
independent
\begin{eqnarray}
\zeta_x &=& \varphi_x+C^{zy}A^{-1}_{yx}\pi_z \nonumber \\
\pi_x &=& \pi_x
\end{eqnarray}
The Jacobean of this transformation is 1 and hence
\begin{equation}
\cD\varphi\cD\pi=\cD\zeta\cD\pi \, .
\end{equation}
The probability distribution (23) in terms of the new variables
$\zeta$ and $\pi$ is
\begin{equation}
P(\zeta,\pi)={1\over W}\exp\left\{-{1\over2}\left(\zeta_xA^{xy}\zeta_y+\pi_x
\Gamma^{xy}\pi_y\right)\right\} \, ,
\end{equation}
where
\begin{equation}
\Gamma^{xy}=B^{xy}-C^{xu}A^{-1}_{uv}C^{vy} \, .
\end{equation}
At this point, the correlation functions can be immediately expressed in terms
of the kernels
\begin{eqnarray}
\vev{\zeta_x\zeta_y} &=& A^{-1}_{xy} \nonumber \\
\vev{\pi_x\pi_y} &=& \Gamma^{-1}_{xy} \nonumber \\
\vev{\zeta_x\pi_y} &=& 0  \, .
\end{eqnarray}
This can be seen either by direct functional integration, or by calculating the
generating functional (characteristic functional)
\begin{equation}
\Phi (J^\zeta,J^\pi)=\int P(\zeta,\pi)\exp\left(-iJ^\zeta_x \zeta^x-iJ^\pi_x
\pi^x\right) \cD\zeta\cD\pi
\end{equation}
and taking second derivatives of it, \eg
\begin{equation}
\vev{\zeta_x\zeta_y}>=-{\delta^2\Phi\over\delta J^\zeta_x\delta J^\zeta_y}
|_{J^\zeta=J^\pi=0}\, .
\end{equation}
Substituting $\zeta$ from (28) into (32) and solving the resulting set of
equations we obtain
\begin{eqnarray}
A^{-1}_{xy} &=& \vev{\varphi_x\varphi_y} - \vev{\varphi_x\pi_u}
\vev{\pi_u\pi_v}^{-1} \vev{\pi_v\varphi_y} \nonumber \\
B^{-1}_{xy} &=& \vev{\pi_x\pi_y} - \vev{\pi_x\varphi_u}
\vev{\varphi_u\varphi_v}^{-1} \vev{\varphi_v\pi_y} \nonumber \\
C^{-1}_{xy} &=& \vev{\varphi_x\pi_y} - \vev{\pi_x\pi_u}
\vev{\varphi_u\pi_v}^{-1} \vev{\varphi_v\varphi_y} \, .
\end{eqnarray}
Deriving the above formulas we took into account that for a spatially
homogeneous Gaussian process the correlation functions depend only on
the difference of the arguments.

To calculate the entropy, we substitute (30) into the general formula (22)
using the fact that the Jacobean of the transformation (28) is unity
and obtain
\begin{eqnarray}
S &=& -\int P(\zeta,\pi) \ln P(\zeta,\pi)\cD\zeta\cD\pi \nonumber \\
&=& V \delta^3(0) + \ln W \nonumber \\
&=& V \delta^3(0) +
\ln\int\exp\left\{-{1\over2}\left(\zeta_xA^{xy}\zeta_y+\pi_x\Gamma^{xy}
\pi_{xy}\right)\right\}\cD\zeta\cD\pi \nonumber \\
&=& V \delta^3(0) + {1\over2}\ln\det (A^{-1}\Gamma^{-1})
\end{eqnarray}
where $V$ is the volume of space.
Dropping the irrelevant constant contributions to the entropy and
inserting  (32) and (35) in the above, we get the following expression for
the entropy in terms of correlation functions
\begin{equation}
S = {1\over2}\ln\det\left(\vev{\varphi_x\varphi_z}\vev{\pi_z\pi_y} -
\vev{\varphi_x\pi_z} \vev{\pi_z\varphi_y}\right) = {1\over 2} \ln \det \cD^{xy}
\end{equation}

Thus, the problem of calculating the entropy has been reduced to the evaluation
of a determinant of the operator
\begin{equation}
\cD^{xy} = \cD (\underline x-\underline y)=
\int(\vev{\varphi(\underline x)\varphi(\underline z)}
\vev{\pi(\underline z)\pi(\underline y)} - \vev{\varphi(\underline x)\pi
(\underline
z)}\vev{\pi(\underline
z)\varphi(\underline y)})d^3z
\, .
\end{equation}
This determinant can be calculated by $\zeta$ function regularization
(see Appendix A). The
result is
\begin{equation}
\det \cD \> \propto \> \exp\left[V\int d^3 k\ln \cD_{\underline k}\right]
\end{equation}
where
\begin{equation}
\cD_{\underline k}={1\over(2\pi)^3} \int d^3\underline x e^{-i\underline
k\cdot\underline x} \cD(\underline x)
\end{equation}
is the spectral density of the operator $\cD(\underline x)= \cD
(\underline{x} - \underline{y})$. Substituting
(39) in (37) we obtain the following expression for the entropy per unit
volume
\begin{eqnarray}
s = {S\over V} &=& \int d^3\underline k\ln\cD_{\underline k} \nonumber
\\
&=& \int d^3\underline {k}\ln\left(\vev{|\varphi_{\underline k}|^2}
 \vev{|\pi_{\underline k}|^2}-\vev{|\varphi_{\underline k}|^2
|\pi_{\underline k}|^2}\right) \, .
\end{eqnarray}
where we expressed the entropy in terms of the spectral density of
correlation functions and omitted irrelevant contributions which do
not depend on the spectrum of the scalar field.

\section{GENERAL COMMENTS}

In this section we shall focus on two issues: the connection between
the two definitions of entropy given in Sections 3 and 4, and a
further discussion of the origin of entropy.

To concretize the analysis we consider a quantized free scalar field
$\hat \varphi (\underline{x}, \eta)$ with time dependent effective
mass.  The field is assumed to start out in its initial vacuum state.
As is well known, this state evolves into a squeezed state [19,20].
Such a state is highly excited in the sense that the expectation value
of the number operator is large.

If the mass is constant at the beginning and at the end, then the
notion of particles is well defined for the in-state and for the out-state,
and in the corresponding time intervals the field operator $\hat
\varphi$ can be expanded either in terms of in-creation and
annihilation operators $\hat a^+$ and $\hat a^-$ or in terms of the
corresponding out-operators $\hat c^+$ and $\hat c^-$:
\begin{eqnarray}
\hat \varphi (\underline{x} , \eta) &=& \int {d^3 k\over{(2
\pi)^{3/2}}} \, \left( e^{i \underline{k} \cdot \underline{x}}
u^{in}_{\underline{k}} (\eta)^\ast \hat a^-_{\underline{k}} + e^{-i
\underline{k} \cdot \underline{x}} u^{in}_{\underline{k}} (\eta) \hat
a^+_{\underline{k}} \right) \nonumber \\
&=& \int {d^3 k\over{(2 \pi)^{3/2}}} \, \left( e^{i \underline{k} \cdot
\underline{x}} u^{out}_{\underline{k}} (\eta)^\ast \hat c^-
_{\underline{k}} + e^{-i \underline{k} \cdot \underline{x}}
u^{out}_{\underline{k}} (\eta) \hat c^+_{\underline{k}} \right)
\end{eqnarray}
where $u_{\underline{k}}^{in} (\eta)$ and $u_\underk^{out} (\eta)$ are
the positive frequency mode functions in the in- and out-states
respectively.  For the particular case under consideration, the
operators $\hat c^+_\underk$ and $\hat c^-_\underk$ are related to
$\hat a^+_\underk$ and $\hat a^-_\underk$ via a Bogoliubov
transformation
\begin{eqnarray}
 c^-_\underk &=& a^-_\underk ch r_k - a^-_{-\underk} e^{2 i
\varphi_k} sh r_k \nonumber \\
c^+_\underk &=& - a^+_{- \underk} \, e^{-2 i \varphi_k} sh r_k +
a^+_{\underk}  ch r_k \, .
\end{eqnarray}
The real functions $r_k (\eta)$ and $\varphi_k (\eta)$ are called
squeeze parameter and squeeze angle respectively [21,22].  The initial
state is taken to be the vacuum $|0 >_{in}$ defined by $\hat a^-
_k | 0 >_{in} = 0$ for all $\underk$.

The expectation value of the number of particles at late times in the
$\underk$ mode is
\begin{equation}
< n_\underk > = < 0_{in} | c_\underk^+ c^-_\underk | 0_{in} > = sh^2
r_k \, .
\end{equation}
Hence, for large values of the squeeze parameter $r_k$, we should
expect that the quantum field $\hat \varphi$ can with good accuracy be
described as a classical field $\varphi_{cl}$, since if $r_k \gg 1$,
the condition $< n_\underk > \gg 1$ to be in the region of
applicability of the classical limit is satisfied.  In this limit, the
correlation functions of the classical field $\varphi_{cl}$ should
coincide with the corresponding expectation values of the operator
$\hat \varphi$, \eg
\begin{equation}
< \varphi_{cl} (\underx , \eta) \varphi_{cl} (\undery , \eta) > \simeq
< 0_{in} | \hat \varphi (\underx , \eta) \hat \varphi (\undery , \eta)
| 0_{in} > \equiv < \hat \varphi (\underx , \eta) \hat \varphi
(\undery , \eta) > \, .
\end{equation}
Hence, in the classical limit both of our formulas for the entropy
should be applicable and give the same result.  In the following we
show that this is indeed true.

First of all, we need to calculate the correlation functions of $\hat
\varphi$ in terms of the squeeze parameters.  Taking into account that
at late times (in the out-state)
\begin{equation}
u^{out}_{\underline{k}} (\eta) \sim e^{i \omega_k \eta} \> , \> \omega_k =
\sqrt{k^2 + m^2_{out} }
\end{equation}
where $m_{out}$ is the mass in the out-phase, and using (43), we find
the following expressions for the late time correlation functions [23]
\begin{eqnarray}
< \varphi (\underx , \eta) \varphi (\undery, \eta) > &=& \int {d^3
k\over{(2 \pi)^3}} \, e^{i \underk \cdot (\underx - \undery)}
{1\over{2 \omega_k}} [ 2 sh^2 r_k + 1 - sh 2 r_k \cos 2 \delta_k]
\nonumber \\
< \pi (\underx , \eta) \pi (\undery , \eta) > &=& \int {d^3 k\over{(2
\pi)^3}} \, e^{i \underk \cdot (\underx - \undery)} {\omega_k\over 2}
\, [ 2 sh^2 r_k + 1 + sh 2 r_k \cos 2 \delta_k ] \nonumber \\
< \varphi (\underx , \eta) \pi (\undery , \eta) > &=& \int {d^3
k\over{(2 \pi)^3}} \, e^{i \underk \cdot (\underx - \undery)} \,
{i\over 2} \, [ 1 - i sh 2r_k \sin 2 \delta_k ] \, ,
\end{eqnarray}
where the angle $\delta_k$ is
\begin{equation}
\delta_k (\eta) = \int d \eta (\omega_k (\eta) - \varphi_k) \, .
\end{equation}
Hence, it follows immediately that
\begin{eqnarray}
\int d^3 z &<& \varphi (\underx) \varphi (\underz) > < \pi (\underz) \pi
(\undery) > \nonumber \\
&=& \int {d^3 k \over{(2 \pi)^3}} \, e^{i \underk \cdot (\underx -
\undery)} {1\over 4} \, [ (2sh^2 r_k +1)^2 - sh^2 2r_k \cos^2 2
\delta_k ]
\end{eqnarray}
and
\begin{eqnarray}
\int d^3 z &<& \varphi (\underx) \pi (\underz) > < \pi (\underz)
\varphi
(\undery) > \nonumber \\
&=& \int {d^3 k \over{(2 \pi)^3}} \, e^{i \underk \cdot (\underx -
\undery)} {1\over 4} \, [ (2sh^2 r_k +1)^2 - sh^2 2r_k \cos^2 2
\delta_k ]
\end{eqnarray}

Taking into account (45) we can substitute (49) and (50) into
(38) to determine the entropy of the field in the classical limit.
The two terms (49) and (50) evidently cancel each other, giving
vanishing entropy.  But this is no surprise since we started in a pure
state whose entropy must vanish and since the evolution of the system
is unitary, thus preserving the entropy.  The information about the
final state is complete.

To associate entropy with the final state we must neglect some
information.  Typically, this will be information which is very
sensitive to any kind of perturbation, either of the system or of the
state.  In our example, the phases $\delta_k$ will depend sensitively
on a perturbation, whereas the amplitudes will not.  In the language
of Section 3, the amplitude is the principal quantum number and the
phases are averaged over.

The nature of the perturbation mentioned above is a subject of
independent analysis.  There are several mechanisms; which one is
realized will depend on the particular system under investigation.
Interactions with other fields (which can be viewed as an environment)
will induce stochasticity of the phases.  Weak self interactions of
the scalar field may also induce large changes in phases.
Approximating the state of the system (\eg neglecting decaying modes
in examples in which the mass is constant at early and late times but
changes in between) will have the same effect.

Returning to our example, let us define the coarse grained entropy by
at first averaging the two point correlation functions (47) over the
phases $\delta_k$ and substituting the thus obtained ``reduced"
correlation functions into (38) to find the reduced operator
$\cD_{red}$ which will be
\begin{equation}
\cD (\underx - \undery) = \int {d^3 k\over{(2 \pi)^3}} sh^2 r_k (1 +
sh^2 r_k) e^{i \underk \cdot (\underx - \undery)} \, .
\end{equation}
Then, according to (41), the classical field definition of entropy
gives
\begin{equation}
s \cong \int d^3 k \ln sh^2 r_k
\end{equation}
in the classical limit $sh^2 r_k \gg 1$.  Taking into account (44) we
obtain
\begin{equation}
s \simeq \int d^3 k \ln n_k
\end{equation}
for $n_k \gg 1$, in agreement with the result for the entropy obtained
in Section 3 (see (19)).

We conclude that in the cases in which both of our formulas for the
entropy from Sections 3 and 4 are applicable, the results coincide as
they should.

\section{COSMOLOGICAL PERTURBATIONS}

Our goal is to apply the definition of nonequilibrium entropy developed in the
previous sections to the gravitational field. Specifically, we will calculate
the entropy of a stochastic background of gravitational waves and of linearized
density perturbations. To set the stage, we briefly review the theory of
linearized cosmological perturbations [24].

We consider linearized perturbations of metric and matter fields about a
homogeneous and isotropic background cosmological model. There are three types
of fluctuations---scalar, vector and tensor perturbations---which are
distinguished by their transformation properties under background space
coordinate changes. Vector modes decay and are irrelevant for cosmology. Hence,
we shall focus on scalar modes (density perturbations) which couple to density
and pressure and tensor modes (gravitational waves).

Although it may at a first glance seem that both density perturbations and
gravitational waves are described by several independent fields, the analysis
can in both cases be reduced to the theory of  a single scalar
field [15,16]; and hence the formalism developed in previous sections to
determine the entropy of a nonequilibrium dynamical system becomes applicable.

\subsection{Gravitational Waves}

Gravitational waves are linearized purely gravitational fluctuations about a
homogeneous and isotropic background metric $g_{\mu \nu}^{(0)}$.  For
simplicity we will consider a spatially flat Universe given by the
invariant line element
\begin{equation}
ds^2=a^2(\eta)(d\eta^2-
\delta_{ij}dx^idx^j) \, ,
\end{equation}
$\eta$ being conformal time and $a(\eta)$ denoting the scale factor.
 The metric
perturbation $\delta g_\munu=a^2h_\munu$ for gravitational
waves is transverse and traceless and satisfies $h_{00}=h_{0i}=0$.

The action $S_{gr}$ for gravitational waves can be obtained by expanding the
Einstein action
\begin{equation}
S=-{1\over6l^2}\int R\sqrt{-
g}d^4x \ \ ,\ \ l^2= {8 \pi
G\over3}
\end{equation}
to second order in the perturbation variables, with the result
\begin{equation}
S_{gr}={1\over24l^2}\int d^4xa^2\left[h^{i^\prime}_{k}
h^{k^\prime}_{i}
-h^i_{k,e}h^k_{i,e}\right],
\end{equation}
where the derivative with respect to $\eta$ is denoted by a prime. Varying this
action yields the following equation of motion for $h^i_j$:
\begin{equation}
h^i_{j^{\prime\prime}} + 2{a'\over a}h^i_{j^{\prime}} - \Delta h^i_j=0 \, .
\end{equation}

In order to reduce the study of gravitational waves to the analysis of a
single scalar field, we expand $h^i_j$ in a Fourier series [15]
\begin{equation}
h^i_j(\underx, \eta)=\int {d^3k\over (2\pi)^3} e^{i\underk\cdot\underx}
G^i_j(\underk)a^{-1} u_k(\eta)
\, ,
\end{equation}
where $G^i_j(\underk)$ is the polarization tensor of a gravitational wave with
wave vector $\underk$. The mode functions $u_k(\eta)$ can be used to define a
scalar field $\varphi$
\begin{equation}
\varphi(\underx,\eta)=({1\over12 \ell^2})^{1/2}\int {d^3k\over(2\pi)^3}
e^{i\underk\cdot\underx} \left(G^i_j(\underk)G^j_i(\underk)\right)^{1/2}
u_k(\eta)
\end{equation}
in terms of which the action (56) becomes
\begin{equation}
S_{gr}={1\over2} \int d^4x\left[\varphi^{\prime^2}- \varphi_{,i}
\varphi_{,i}
+{a^{\prime\prime}
\over a} \varphi^2\right] \, .
\end{equation}
This action coincides with the action of a free scalar field with time
dependent mass
$m(m^2=-a^{\prime\prime}/a)$ in flat space-time.

Based on the above equivalence between the gravitational radiation field and
the scalar field $\varphi (\underx,\eta)$, we can use the general definitions
of
entropy developed in Sections 3 and 4 to define the entropy in gravitational
radiation. This will be done in Section 7.

\subsection{Density Perturbations}

The theory of linearized density perturbations can also be reduced to the
analysis of a single scalar field [25].  However, the reduction process
is more complicated than in the case of gravitational waves.

Density perturbations are scalar type metric fluctuations which couple to
energy density and pressure. At first sight, the most general scalar type
metric perturbation $\delta g_\munu(\underx,\eta)$
 can be expressed in terms of four
free functions. However, two of these functions describe pure gauge
modes [26], \ie inhomogeneities which correspond to a change of the
background space-time coordinates. The easiest and most physical way to avoid
gauge artifacts is to adopt a manifestly gauge invariant formalism
[26]
(for a pedagogical introduction see Ref.~16, for a comprehensive review see
Ref.~15).
In this approach, scalar type metric perturbations are characterized by two
functions which are, via the linearized Einstein equations, coupled to matter
inhomogeneities. The two gauge invariant functions $\Phi(\underx,\eta)$ and
$\Psi(\underx,\eta)$ can easily be identified by transforming to longitudinal
gauge (the system of coordinates in which $\delta g_\munu$ is diagonal):
\begin{equation}
\delta
g_\munu=a^2\pmatrix{2\Phi&0\cr
0&2\Psi\delta_{ij}}
\end{equation}
For scalar matter and for an ideal gas, $\Phi=\Psi$ as a consequence of the
$i\neq j$ Einstein equation (for these forms of matter $\delta T_{ij}\sim
\delta_{ij}$).

At this point, the description of density perturbations has been reduced to
prescribing a gauge invariant combination $v$ of matter perturbation
and metric fluctuations in terms of which the action for
perturbations can be expressed in the form [27]:
\begin{equation}
S_v={1\over 2}\int d^4 x\left[v^{\prime^2} - c^2_s v_{,i} v_{,j}
\delta^{ij} +
{z^{\prime\prime}\over z}v^2\right]
\end{equation}
where $c_s^2=1$ for scalar field matter and $c_s^2=p/\rho$ for ideal gas
matter. The variable $z(\eta)$ is a combination of background dependent
factors. For a perfect fluid and for a scalar field as matter
\begin{equation}
z={a\beta^{1/2}\over \cH
c_s}~~,~~~\beta=\cH^2-
\cH^\prime \, ,
\end{equation}
where $\cH = a^\prime / a$.

All gauge invariant variables (like $\Phi$) can be expressed in terms
of the variable $v$ via the Einstein equations.  For example, the
metric potential $\Phi$ which characterizes the amplitude of metric
perturbations in longitudinal gauge (see (61)) is expressed in terms
of $v$ in the following manner
\begin{equation}
\Delta \Phi = - \sqrt{{3\over 2}} \ell {\beta\over{\cH c_s^2}} \,
\left({v\over z}\right)^\prime
\end{equation}
(valid for both scalar field and hydrodynamical matter).

In conclusion, we have been able to reduce the action for scalar metric
perturbations to that of a free scalar field with mass
$m^2=-z^{\prime\prime}/z$.
However, in contrast to the case of gravitational perturbations, the spatial
gradient term has a prefactor $c_s^2$. In the radiation dominated
period of the evolution of the Universe this term implies that density
perturbations propagate with the speed of sound $c_s=1/\sqrt{3}$. In the matter
dominated period, $c_s = 0$. This has important consequences for
our ability to define the notion of ``particle number" of a given
field configuration for different equations state.

\section{ENTROPY OF COSMOLOGICAL PERTURBATIONS}

In the previous section we demonstrated that the action for both gravitational
waves and for density perturbations can be reduced to the action of a free
scalar field. Hence the formalism of sections 3 and 4 can be used to define the
nonequilibrium entropy of the gravitational field.

In order to use the quantum definition of entropy of Section 3 we need
a well defined notion of particles.
For gravitational waves described by the action
(60) and in the special case when
$a^{\prime\prime}=0$ (which is realized in the radiation dominated period when
$a(\eta)\sim\eta$) the mass term in (60) vanishes, and hence the
notion of particles is well defined and the particle number remains constant.
In
contrast, during an inflationary period $a(\eta)=-1/(H\eta)$, and
this leads to
a negative time dependent effective square mass $m^2 = -
{a^{\prime\prime}\over a}$ in (60) which is
\begin{equation}
m^2 =-{2\over\eta^2}
\end{equation}
Because of this time dependent mass, there will be in this case
particle production and corresponding increase in entropy.  Focusing
on
modes with comoving wavenumber $k$, the time dependent mass induces an
effective potential
\begin{equation}
V_{eff,k}(\varphi)={1\over2}\left(k^2-{2\over\eta^2}\right)\varphi^2
\, .
\end{equation}
Hence, the increase in particle number is significant on scales which
satisfy
\begin{equation}
k<-{\sqrt{2}\over\eta}=\sqrt{2} Ha,
\end{equation}
which (up to a factor of $\sqrt{2}$) is exactly the condition for the
wavelength
$k^{-1}a$ to be larger than the Hubble radius $H^{-1}$. Similarly, in the
matter dominated period when $a(\eta)\sim\eta^2$ there is a nonvanishing
effective potential. In terms of $\eta$ the potential takes precisely the form
(65). Hence, (66) and (67) also apply in this case.

In the following, we shall evaluate the entropy for the spectrum of
gravitational waves produced in an inflationary Universe. From the previous
discussion, we see that the magnitude of the entropy is generated by
``parametric amplification" [28] of the number of gravitons in each mode
which occurs during inflation on scales larger than the Hubble radius.
 On scales which enter the Hubble radius after $t_{eq}$, the time of equal
matter and radiation, an additional amplification takes place during the matter
dominated epoch. However, we emphasize that although the magnitude of the
``potential" entropy is set by the inflationary phase, in order to get any
nonvanishing entropy there needs to be a loss of correlations due to some
coarse graining (see Section 5).

The situation for density perturbations is similar.  During the
period of radiation domination, it follows from (6.12) that
$z^{\prime\prime}=0$. Hence, as in the case of gravitational waves,
the notion of particles is well defined, the particle number is time
independent, and hence
the
entropy per mode remains constant. During inflation, it follows from
(63)
that $z^{\prime\prime}/z\simeq a^{\prime\prime}/a=2/\eta^2$, and that therefore
like for gravitational waves particle creation for modes with wavelengths
larger
than the sound horizon $H^{-1}$ occurs, which leads to an increase in
the occupation number of each mode.  The analysis of density
perturbations is more complex in the matter dominated period since
$c_s^2=0$. Hence no occupation
number can be defined. In this case we must use the classical framework of
Section 4 in order to define the entropy.

In the following, we will first calculate the entropy per mode of gravitational
waves and density perturbations which were produced during inflation
in a cosmological model in which the Universe is radiation dominated
at late times and argue that the result for the more realistic case is
the same.

\subsection{Gravitational Radiation}

To estimate the entropy of gravitational radiation produced during
inflation, it is convenient to express the expectation value $<
n_\underk >$ of the number operator in each mode $\underk$ (or,
correspondingly, the squeeze parameter $r_k$) in terms of the spectral
density of the two point correlation function of the metric $h^i_j$.
Since this spectrum $| \delta^h (k) |$ was calculated in the
particular models we are interested in [29], we can then use these
results to estimate the entropy via formulas (52) or (53) (the
result will be the same since on the scales of interest $n_k \gg 1)$.

We can only justify the appearance of entropy for perturbations on
scales smaller than the Hubble radius, since the mode functions do not
oscillate on larger scales.  Hence, when calculating the entropy in
gravitational radiation at some late time, we will only consider
scales smaller than the Hubble radius at that time.

Combining (58) and
(59) with the standard mode expansion of the quantum operator
$\hat\varphi$
associated with $\varphi$, we obtain the following result for the operator
$\hat
h^i_j$:
\begin{equation}
\hat h^i_j (\underx,\eta) = (6l^2)^{1/2} a^{-1} \int {d^3k\over(2\pi)^{3/2}}
k^{-1/2}   {G^i_j(\underk)\over(G^m_n(\underk)G^n_m(\underk))^{1/2}}
\left[e^{-i \underk \cdot \underx} u_\underk^{out} (\eta)^\ast c_\underk
+ e^{i \underk \cdot \underx} u_\underk^{out} (\eta)
c_\underk^+\right]
\end{equation}
where $c^+_\underk$ and $c_\underk^-$ are creation and annihilation operators
respectively of particles with comoving wave vector $\underk$ at some
late time and $u^{out}_\underk (\eta) \simeq e^{i k \eta}$ for
perturbations on scales smaller than the Hubble radius $(k \eta \gg
1)$.

The spectrum $\delta_h(\underk)$ of gravitational radiation is defined in terms
of the two point function of $\hat h^i_j$ by
\begin{equation}
\vev{0_{in} |\hat h^i_j(\underx,\eta)
\hat h^j_i(\underx + \underline{r},\eta)|0_{in}} =
\int^\infty_0 {dk\over k} {\sin(kr)\over kr} |\delta_h(k)|^2
\end{equation}
with $r=(\underr)$ and $k=|\underk|$ and where $|0_{in} >$ is the initial
vacuum state defined by $a^-_\underk |0_{in} > = 0$.  On the other hand,
this two point function can be evaluated in terms of occupation numbers
using (68)
\begin{equation}
\vev{\psi|\hat h^i_j(\underx,\eta)\hat h^j_i(\underx + \underline{r},\eta)|0}
= {6l^2\over2\pi}  a^{-2} \int^\infty_0 {dk\over k} {\sin kr\over kr} k^2
(2 <n_k> +1)
\end{equation}
where we took into account that $<0_{in}| \hat c_k^+ \hat c_k^- |
0_{in} > = <n_k>$.
Comparing (69) and (70) yields
\begin{equation}
2 <n_k> +1 = {2\pi\over6l^2}
{|\delta_h(k)|^2\over(k/a)^2}
\, .
\end{equation}

Equation (71) can be applied to calculate the entropy of gravitational
radiation in any cosmological model. Given the spectrum $\delta_h(k)$, the
occupation numbers $<n_k>$ are determined for each mode by (71), and the
entropy
per mode follows from (53).

To be specific, we evaluate the entropy in an inflationary Universe. The
spectrum of gravitational radiation originating in quantum fluctuations during
the phase of exponential expansion has been calculated many times
[29] (for an explicit derivation using the notation of this paper see
Ref.~15). The
result is
\begin{equation}
\delta_h(k) = {\ell H\over\sqrt{2}\pi}
\left\{ \begin{array}{ll}
(t_0k)^{-2} &  t_0^{-1} <k<t_0^{-1} z^{1/2}_{eq} \\
(t_0k)^{-1} z_{eq}^{-1/2} \,\, & z_{eq}^{1/2} t_0^{-1}<k<t_0^{-1}
                   z_{eq}^{-1/4} ({t_0\over t_R})^{-1/2} \; ,
\end{array}\right.  \\
\end{equation}
where $t_0$ is the present time and $t_R$ corresponds to the end of inflation.
$z_{eq}$ is the redshift at the time of equal matter and radiation. The top
line corresponds to scales which enter the Hubble radius after $t_{eq}$, the
bottom to those which enter during the period of radiation domination. Hence,
from Equation (71)
\begin{equation}
2 <n_k> +1 = {H^2\over6\pi k^2} \left\{
\begin{array}{ll}
(t_0k)^{-4} & t_0^{-1}<k<t_0^{-1} z_{eq}^{1/2}  \\
(t_0k)^{-2} z_{eq}^{-1} \,\, & z_{eq}^{1/2}t_0^{-1}<k<t_0^{-1}
z_{eq}^{-1/4} ({t_0\over t_R})^{1/2} \; ,
\end{array}\right. \\
\end{equation}
$H$ being the Hubble expansion rate during the period of inflation.

Choosing $H=10^{13}GeV$, a value for which fluctuations from inflation have the
right order of magnitude to seed galaxies [30], and comparing the entropy
$s_k$ per mode on galactic scales $k^{-1}\sim10Mpc$, we conclude from (73)
that
\begin{equation}
s_k \sim 100\ln 10 \, .
\end{equation}

Formula (74) can also be interpreted as giving the entropy in a
volume $V=k^{-3}$ of gravitational waves with wave number of the order
$k$.  This quantity can be compared with the statistical fluctuations
of the entropy of the cosmic microwave background (CMB).  These
fluctuations $\Delta s_{\mbox{\scriptsize{CMB}}} (k)$ scale as
\begin{equation}
\Delta s_{\mbox{\scriptsize{CMB}}} (k) \sim (k \lambda_\gamma)^{3/2} \, ,
\end{equation}
where $\lambda_\gamma$ is the characteristic wavelength of the CMB.
Hence, for large $k^{-1}$ (such as the scales mentioned above), the
entropy in gravitational waves exceeds that in the statistical
fluctuations of the CMB.  This reflects the result that on
cosmological distance scales the fluctuations produced by inflation
are much larger than the Poisson noise in the background fields.

However, the total entropy density for the gravitational background is smaller
than that of the CMB for which
\begin{equation}
S_{\mbox{\scriptsize{CMB}}}\simeq {4\pi\over3}
T_0^3 \, .
\end{equation}
{}From (53) it follows that the entropy density in gravitational
radiation is
\begin{equation}
S_{gr} \simeq {4\pi\over3} k_c^3
\end{equation}
where $k_c$ is determined by
\begin{equation}
n(k_c) \simeq 1 \, .
\end{equation}
{}From (73) it follows that
\begin{equation}
k_{c} \sim t_0^{-1} z_{eq}^{-1/4} ({t_0\over t_R})^{1/2}
\end{equation}
and hence
\begin{equation}
S_{gw} \sim \left({H\over m_{pl}} \right)^{3/2} T_0^3
\end{equation}
Comparing (77) and (80) we see that the total entropy density in
gravitational waves is suppressed compared to the entropy of the CMB by a
factor $(H/m_{pl})^{3/2}$. Nevertheless, as seen above, on large length
scales, gravitational radiation dominates over the fluctuations of the
entropy of the CMB.

\subsection{Density Perturbations}

The approach for calculating the entropy of density perturbations
follows what was done above for gravitational waves. Starting point is the
expansion of the operator $\hat \varphi$ associated with the scalar field $v$
of
(62) [which contains all the information about density perturbations] in
terms of late time creation and annihilation operators $c_k^+$ and $c^-_k$
respectively:
\begin{equation}
\hat \varphi (\underx, \eta) = {1\over\sqrt{2}} \int {d^3k\over(2\pi)^{3/2}}
\left[e^{i\underk\cdot\underx} v^{out}_k (\eta) c^-_k
+ e^{-i\underk\cdot\underx} v^{out}_k (\eta) c^+_k\right]
\end{equation}

To simplify the consideration and to justify the notion of entropy, we will
estimate the entropy on scales smaller than the Hubble radius at late times,
assuming that in this late time interval the Universe is radiation dominated.
In this case, the notion of particles is unambiguous since the mode functions
$v_k (\eta)$ take the form
\begin{equation}
v_k (\eta) = {1\over{{\sqrt{\omega}}}} e^{i \omega \eta}
\end{equation}
with $\omega=c_s k = k / \sqrt{3}$.

Using the relation (64) between the gauge invariant relativistic potential
$\Phi$ and
the scalar field $v$ valid when $c_s^2\neq0$ we obtain [15]
\begin{equation}
\hat\Phi (\underx, \eta) = ({3\over4})^{1/2} l {\beta^{1/2}\over a} \int
{d^3k\over(2\pi)^{3/2}} \left[ e^{i\underk\cdot\underx} u^*_k(\eta) c^-_k
+ e^{-i\underk\cdot\underx} u_k (\eta) c_k^+\right]
\end{equation}
with
\begin{equation}
u_k={z\over k^2 c_s} ({v_k\over z})^\prime \, .
\end{equation}

As was done previously in the case of gravitational waves, the next step is to
derive the relation between the spectrum $\delta_\Phi(k)$ of density
perturbations
\begin{equation}
\vev{0_{in} | \hat\Phi (0, \eta) \hat\Phi (\underr, \eta) |0_{in} }
= \int^\infty_0 {dk\over k} {\sin kr\over kr} |\delta_\Phi(k)|^2
\end{equation}
and the occupation number $<n_k>$ defined with respect to the creation and
annihilation operators introduced in (81).  Using (83) it follows
that
\begin{equation}
\vev{0_{in} |\hat\Phi(0,\eta)\hat\Phi(\underr,\eta)| 0_{in}} = {3\over4} l^2
{\beta\over a^2} \int {d^3k\over(2\pi)^3} (2 <n_k> +1) u_k^* u_k
e^{i\underk\cdot\underr} \, .
\end{equation}
Making use of (82) and (84), and evaluating $z$ during the period of
radiation domination, we obtain
\begin{eqnarray}
\vev{0_{in}|\hat\Phi(0,\eta)\hat\Phi(\underr,\eta)| 0_{in}} & = &
{9\sqrt{3}\over(2\pi)^2} l^2\eta^{-4} a^{-2} \int^\infty_0 dkk^{-1}
{\sin kr\over kr} \\ \nonumber
& \times & (2 <n_k> +1) k^{-2} \left(1+{(k\eta)^2\over3}\right)
\, .
\end{eqnarray}
Comparing (85) and (87) yields
\begin{equation}
2 <n_k> +1 = {(2\pi)^2\over 9\sqrt{3}} (k\eta)^2 (a\eta)^2 l^{-2}
\left(1+{(k\eta)^2\over3} \right)^{-1}|\delta_\Phi(k)|^2
\, ,
\end{equation}
which expresses the occupation numbers in terms of the spectrum of the
relativistic potential for density perturbations.

The entropy for density perturbations can now be determined by
combining (88)
and (53).  To be specific, we evaluate the entropy per mode in a model of
chaotic inflation [31] with potential
\begin{equation}
V (\varphi) = {1\over2}
m^2\varphi^2 \, .
\end{equation}
In this case, the spectrum of density perturbations immediately after
inflation on scales which are larger than the Hubble radius
is given by [15]
\begin{equation}
|\delta^\Phi_k| \simeq {\sqrt{2}\over{3\pi}} l m \ln
({\lambda\over{\lambda_\gamma}})
\end{equation}
where $\lambda_\gamma$ is the characteristic wavelength of the CMB.
To obtain the late time spectrum, the decay of the amplitude of $\Phi$ on
scales inside the Hubble radius must be taken into account.  Assuming that the
Universe is radiation dominated after inflation we obtain [15]
\begin{equation}
| \delta^\Phi_k | \simeq {\sqrt{2}\over{3 \pi}} \ln \ln \left(
{\lambda\over{\lambda_\gamma}} \right) \, \left( {\lambda\over{t_0}} \right)^2
\end{equation}
for $k \eta > 1$ at time $t_0$.

Inserting (91) into (88) yields the following result for $k \eta \gg 1$:

\begin{equation}
2 <n_k> +1 \sim \left( mt_0 \right)^2 \ln^2 \left( {\lambda\over\lambda_\gamma}
\right) \left({\lambda\over{t_0}} \right)^4 \, .
\end{equation}
Up to the logarithmic factor, this result agrees with the corresponding result
(73) for gravitational waves.

For $m\sim10^{13}$ GeV (the upper bound on $m$ from constraints on the
anisotropy of the CMB) we hence obtain the same entropy density per
mode
\begin{equation}
s_k \sim 100\ln 10 \, .
\end{equation}

We conclude that the entropy per mode of density perturbations on large scales
($\lambda\gg\lambda_\gamma$), in particular on scales of galaxies, clusters and
large-scale structure, exceeds the statistical fluctuations of the
entropy of the CMB. This supports the important
role of the entropy of the gravitational field in the galaxy formation
process.

The quantum approach of calculating the entropy of density perturbations breaks
down during the matter dominated era. However, based on the considerations of
Section 5, it is easy to extend the analysis.

As discussed in Section 4, the classical definition (41) can easily be applied
during the epoch of matter domination. The results obtained above will
not change significantly.  However, there are some interesting
questions concerning coarse graining and time dependence of the
entropy in a matter dominated Universe which we shall consider
elsewhere.

Formulas (71) and (88) can be used to calculate the entropy of
gravitational waves and density perturbations in any cosmological
model in which the spectra $\delta_h(k)$ and $delta_\Phi(k)$ are
known. In particular, the entropy of the gravitational field in
topological defect models of structure formation [14] can easily be
determined.

\section{CONCLUSIONS}

We have presented two general definitions of nonequilibrium entropy and applied
them to calculate the entropy in gravitational radiation and to give a measure
of the entropy of linear density fluctuations
and gravitational waves in a Friedmann Universe.

Our first definition of entropy is based on the microcanonical ensemble and is
applicable to systems with well defined occupation numbers. The origin of the
entropy is coarse graining: ignoring correlations in the form of information
about quantum numbers other than the principal quantum number which is usually
taken to be the energy.

The second definition of entropy applies to any stochastic classical field and
expresses the entropy in terms of two point correlation functions. The physical
origin of entropy in this case is also due to coarse graining.  We have shown
that the entropy of scalar fields in an expanding Universe
satisfies the second law of thermodynamics. Any increase in entropy during
Hamiltonian evolution is a consequence of coarse graining.

We have used the gauge invariant theory of cosmological perturbations to give a
consistent and unified definition of entropy of cosmological perturbations. On
scales of galaxies and larger, this entropy is larger than the
statistical fluctuations in the entropy of CMB
photons on these scales. Hence, this entropy is important for structure
formation in the Universe. However, the total entropy in density perturbations
and in gravitational waves is smaller than the total entropy of the CMB.

It is our hope that the methods presented here can be used in many different
situations. In particular, they might allow a definition of entropy of density
perturbations beyond linear theory.

\centerline{ACKNOWLEDGEMENTS}

For stimulating discussions we thank Andy Albrecht, Leonid Grishchuk,
Jim Hartle, Tony Houghton, Bei-Lok Hu and Henry Kandrup. Two of us
(R.B. and T.P.) thank the ITP of the University of California in Santa
Barbara for hospitality during the completion of this work. V.M.
thanks the Swiss National Science Foundation for financial support. At
Brown, this research was supported by DOE grant DE-AC02-76ER03130 Task A
and by an Alfred P. Sloan Foundation Fellowship to R.B..
Financial support from NSF grant PHY89-04035 at the ITP in Santa
Barbara is acknowledged.

\unletteredappendix{}
\centerline{\underbar{Calculation of the Determinant of
$\cD (\underx - \undery) \approx \cD^{xy}$}}

The determinant of the operator $\cD (\underx - \undery) \approx \cD^{xy}$
arising in (38) can be calculated using the $\zeta$-function method [32].
More generally, consider an operator $\cD^{xy}$ with a discrete set of
positive real eigenvalues $a_i$ and eigenfunctions
\begin{eqnarray}
f^{(i)} (\underx) &\approx & f^{(i)}_x \nonumber \\
\cD^{xy} f^{(i)}_y &=& \int d^3y \cD (\underx - \undery) f^{(i)} (\undery) =
a_i \delta^{xy} f^{(i)}_y \, .
\end{eqnarray}
This set of operators includes any operator with positive spectral density
\begin{equation}
\cD (\underk) = \int d^3 ze^{-i \underk \cdot \underz} \cD (\underz)
\end{equation}
which has support in a finite volume $V$, \ie $\cD (\underz) = 0$ if $\underz
\notin V$.

The $\zeta$-function associated with such an operator is
\begin{equation}
\zeta_\cD (s) = \sum_i \, {1\over{a^s_i}} \, ,
\end{equation}
where the sum extends over all nonzero eigenvalues.  It follows that
\begin{equation}
\zeta^\prime_\cD (0) = {{d \zeta_\cD(s)}\over{ds}} |_{x=0}
= - \sum_i \ln a_i e^{-s a_i} |_{s=0} = - \ln \left(\Pi_i a_i \right)
\end{equation}
and hence
\begin{equation}
\det \cD = \Pi_i a_i = e^{- \zeta^\prime_\cD (0)} \, .
\end{equation}
Thus, the calculation of the determinant of $\cD$ has been reduced to the
evaluation of $\zeta^\prime_\cD (0)$.

Let us now introduce the heat kernal $G (\underx , \undery, \tau)$ associated
with $\cD$,
\begin{equation}
G (\underx , \undery, \tau) = \sum_i e^{-a_i \tau} f^{(i)}_x
f^{(i)^\ast}_y \, ,
\end{equation}
which in the case under consideration depends only on $\underx - \undery$ and
$\tau$ and satisfies the equation
\begin{equation}
\int \cD (\underx - \underz) G (\underz - \undery) d^3 \underz = - {\partial G
(\underx - \undery, \tau)\over{\partial \tau}}
\end{equation}
with boundary condition
\begin{equation}
G(\underx - \undery, 0) = \delta (\underx - \undery) \, ,
\end{equation}
a consequence of the completeness of the set of eigenfunctions.  It is easy to
check by explicit integration that
\begin{equation}
\zeta_\cD (s) = {1\over{\Gamma (s)}} \int^\infty_0 d \tau \tau^{s-1} \int d^3
\underx G (\underx , \underx, \tau) \, .
\end{equation}

Thus, in order to compute $\det \cD$ we need to solve Eq. (A7) for $G(\underx
, \undery, \tau)$ given the boundary condition (A8).  It is convenient to work
in Fourier space where Eq. (A7) takes the form
\begin{equation}
\cD (\underk) G (\underk , \tau) = - {\partial\over{\partial  \tau}} (G
(\underk, \tau ))
\end{equation}
with boundary condition
\begin{equation}
G (\underk , 0) = 1 \, ,
\end{equation}
where
\begin{equation}
G (\underk , \tau) = \int G (\underz , \tau) e^{i \underk \cdot
\underz} d^3 z \, .
\end{equation}
As a result we obtain the following solution of (A7):
\begin{equation}
G (\underx - \undery, \tau) = \int d^3 \underk e^{\cD (\underk) \tau}
e^{i \underk (\underx - \undery)} \, .
\end{equation}
Substituting (A13) into (A9) on finds
\begin{equation}
\zeta_{\cD} (s) = V \int d^3 \underk e^{-s \ln \cD (\underk)}
\end{equation}
and correspondingly
\begin{equation}
\det \cD = \exp \left( V \int \ln \cD (\underk) d^3 \underk \right)
\end{equation}
where $V$ is the volume of the support of $\cD$.


\begin{references}
\bibitem[*]{slava}On leave of absence from Institute for Nuclear
Research, Academy of Sciences, Moscow, Russia.
\bibitem{1} J. Bekenstein, {\it Phys. Rev.} {\bf D7}, 2333 (1973).
\bibitem{2} S. Hawking, {\it  Comm. Math. Phys.} {\bf 43}, 199 (1975).
\bibitem{3} R. Penrose, in `General Relativity: An Einstein Centenary Survey'
ed. by S. Hawking and W. Israel (Cambridge Univ. Press, Cambridge, 1979).
\bibitem{4.} C. Shannon, {\it Bell System Tech. J.} {\bf 27}, 379, 623
(1948);\\
E. Jaynes, {\it Phys. Rev.} {\bf 106}, 620 (1957).
\bibitem{5} L. Smolin, {\it Gen. Rel. Grav.} {\bf 17}, 417 (1985).
\bibitem{6}  B. Hu and D. Pavon, {\it Phys. Lett.} {\bf 180B}, 329 (1986).
\bibitem{7} H. Kandrup, {\it  J. Math. Phys.} {\bf 28}, 1398 (1987).
\bibitem{8} H. Kandrup, {\it Phys. Rev.} {\bf D37}, 3505 (1988).
\bibitem{9} B. Hu and H. Kandrup, {\it Phys. Rev.} {\bf D35}, 1776 (1987).
\bibitem{10} S. Habib and H. Kandrup, {\it  Ann. Phys.} {\bf 191}, 335
(1989).
\bibitem{11} B. Hu, {\it Phys. Lett.} {\bf 97A}, 368 (1983).
\bibitem{12} K. Huang, `Statistical Mechanics' (Wiley, New York, 1963);\\
R. Kubo, `Statistical Mechanics' (Elsevier, Amsterdam, 1965);\\
L. Landau and E. Lifshitz, `Statistical Physics' (Pergamon Press, London,
1959).
\bibitem{13} A. Guth, {\it Phys. Rev.} {\bf D23}, 347 (1981).
\bibitem{14} T.W.B. Kibble, {\it  J. Phys.} {\bf A9}, 1387 (1976);\\
A. Vilenkin, {\it  Phys. Rep.} {\bf 121}, 263 (1985).
\bibitem{15} V. Mukhanov, H. Feldman and R. Brandenberger,
{\it Phys.\ Rep.} {\bf 215}, 203 (1992).
\bibitem{16} R. Brandenberger, H. Feldman and V. Mukhanov, `Gauge Invariant
Cosmological Perturbations', Brown preprint BROWN-HET-845 (1992), to be
published in the proceedings of ICGC-91 (Wiley Eastern Ltd. , New Delhi, 1992).
\bibitem{17} V. Mukhanov, {\it Pisma\ Zh.\ Eksp.\ Fiz.\/} {\bf 44\/}, 50
(1986).
\bibitem{18} E. Calzetta and B. Hu, {\it Phys. Rev.} {\bf D37}, 2878
(1988).
\bibitem{19} L. Grishchuk and Yu. Sidorov, {\it Class. Quant. Grav.}
{\bf 6}, L161 (1989);\\
L. Grishchuk and Yu. Sidorov, {\it Phys. Rev.} {\bf D42}, 3413 (1990).
\bibitem{20} L. Grishchuk, `Quantum mechanics of the primordial cosmological
perturbations', to be published in the proceedings of the 6th Marcel Grossmann
meeting, Kyoto, 1991.
\bibitem{21} C. Caves, \etal\ {\it  Rev. Mod. Phys.} {\bf 52}, 341 (1980).
\bibitem{22} B. Schumaker, {\it  Phys. Rep.} {\bf 135}, 317 (1986).
\bibitem{23} T. Prokopec, Brown Univ. preprint BROWN-HET-861 (1992).
\bibitem{24} see \eg\ S. Weinberg, `Gravitation and Cosmology' (Wiley, New
York,
1972); \\
P.J.E. Peebles, `The Large-Scale Structure of the Universe' (Princeton
Univ. Press, Princeton, 1980);\\
Ya. B. Zel'dovich and I. Novikov, `The Structure
and Evolution of the Universe' (Chicago Univ. Press, Chicago, 1983).
\bibitem{25} G. Chibisov and V. Mukhanov, `Theory of Relativistic Potential:
Cosmological Perturbations', preprint No. 154 of P.N. Lebedev Physical
Institute (1983).
\bibitem{26} U. Gerlach and U. Sengupta, {\it Phys. Rev.} {\bf D18},
1789 (1978);\\ J. Bardeen, {\it Phys. Rev.} {\bf D22}, 1882 (1980).
\bibitem{27} V. Mukhanov, {\it  Zh. Eksp. Teor. Fiz.} {\bf 94}, 1 (1988).
\bibitem{28} L. Grishchuk, {\it  Zh. Eksp. Teor. Fiz.} {\bf 67}, 825
(1974).
\bibitem{29} A. Starobinsky, {\it Pisma~~Zh.~Eksp.\ Teor.\ Fiz.\/} {\bf 30},
719 (1979);\\
V. Rubakov, M. Sazhin and A. Veryaskin, {\it Phys. Lett.} {\bf 115B},
189 (1982);\\
V. Mukhanov, PhD thesis, Lebedev Physical Institute (1982).
R. Fabbri and M. Pollock, {\it Phys. Lett.} {\bf 125B}, 445 (1983);\\
L. Abbott and M. Wise, {\it Nucl. Phys.} {\bf B244}, 541 (1984);\\
L. Abbott and D. Harari, {\it Nucl. Phys.} {\bf B264}, 487 (1986);\\
B. Allen, {\it Phys. Rev.} {\bf D37}, 2078 (1988);\\
V. Sahni, {\it Phys. Rev.} {\bf D42}, 453 (1990).
\bibitem{30} J. Bardeen, P. Steinhardt and M. Turner, {\it Phys. Rev.}
{\bf D28}, 679 (1983);\\
R. Brandenberger and R. Kahn, {\it Phys. Rev.} {\bf D29}, 2172 (1984);\\
V. Mukhanov, {\it Pisma\ Zh.\ Teor.\ Fiz.\/} {\bf 41\/}, 402 (1985);\\
D. Lyth, {\it Phys. Rev.} {\bf D31}, 1792 (1985).
\bibitem{31} A. Linde, {\it Phys. Lett.} {\bf 129B}, 177 (1983).
\bibitem{32} see e.g. P. Ramond, `Field Theory: A Modern Primer'
(Benjamin, Reading, 1981).
\end{references}
\end{document}